\documentclass[apj]{emulateapj}								
\usepackage{multirow}
\usepackage{longtable}
\usepackage{rotating}

\def\gtorder{\mathrel{\raise.3ex\hbox{$>$}\mkern-14mu
             \lower0.6ex\hbox{$\sim$}}}
\def\ltorder{\mathrel{\raise.3ex\hbox{$<$}\mkern-14mu
             \lower0.6ex\hbox{$\sim$}}}

\slugcomment{Draft of \today}

\shorttitle{Rotational Resolved Spectroscopy of Asteroid Pairs}

\shortauthors{Polishook et al.}

\begin{document}
\title{Rotationally Resolved Spectroscopy of Asteroid Pairs: No Spectral Variation Suggests Fission is followed by Settling of Dust}
\author{D.~Polishook\altaffilmark{1},
N. Moskovitz \altaffilmark{1,2},
F.~E. DeMeo \altaffilmark{1,3},
R.~P. Binzel \altaffilmark{1}
}

\altaffiltext{1}{Department of Earth, Atmospheric, and Planetary Sciences, Massachusetts Institute of Technology, Cambridge, MA 02139, USA}
\altaffiltext{2}{Lowell Observatory, 1400 West Mars Hill Road, Flagstaff, AZ 86001, USA}
\altaffiltext{3}{Harvard-Smithsonian Center for Astrophysics, 60 Garden Street, Cambridge, MA 02138, USA}

\begin{abstract}
	The fission of an asteroid due to fast rotation can expose sub-surface material that was never previously exposed to any {\it space weathering} process. We examine the spectral properties of {\it asteroid pairs} that were disrupted in the last 2 million years to examine whether the site of the fission can be revealed. We studied the possibility that the sub-surface material, perhaps on one hemisphere, has spectral characteristics differing from the original weathered surface. This was achieved by performing rotationally-resolved spectroscopic observations to look for local variations as the asteroid rotates.
	
	We spectrally observed 11 asteroids in pairs in the near-IR and visible wavelength range. Photometric observations were also conducted to derive the asteroid lightcurves and to determine the rotational phases of the spectral observations. We do not detect any rotational spectral variations within the signal-to-noise of our measurements, which allows us to tightly constrain the extent of any existing surface heterogeneity.
	
	For each observed spectrum of a longitudinal segment of our measured asteroids, we estimate the maximal size of an un-detected ``spot" with a spectral signature different than the average. For five asteroids the maximal diameter of such a ``spot" is smaller by a factor of two than the diameter of the secondary member of the asteroid. Therefore, the site of the fission is larger than any area with a unique spectral parameters. This means the site of the fission does not have a unique spectrum. In the case of an ordinary chondrite asteroid (S-complex), where the site of fission is expected to present non-weathered spectra, a lack of a fission ``spot" (detectable spectroscopically) can be explained if the rotational-fission process is followed by the spread of dust that re-accumulates on the primary asteroid and covers it homogeneously. This is demonstrated for the young asteroid 6070 that presents an Sq-type spectrum while its inner material, that is presumably revealed on the surface of its secondary member, 54827, has a non-weathered, Q-type spectrum. The spread of dust observed in the disintegration event of the asteroid P/2013 R3, might be an example of such a process and an indication that P/2013 R3 was indeed formed in a rotational-fission event.


\end{abstract}

\keywords{
Asteroids; Asteroids, rotation; Asteroids, surfaces; Rotational dynamics; Spectroscopy}

\section{Introduction and Motivation}
\label{sec:Introduction}

	The rotational-fission mechanism can result in mass shedding or complete disintegration of a strengthless {\it rubble-pile} asteroid that fails to remain bound due to an increasing rotation rate. This physical process is backed by multiple independent observations: asteroids were measured to have low density values that reflect high porosity and shattered structure (e.g. (25143) Itokawa; Fujiwara et al. 2006, Carry 2012); asteroids were measured while spinning-up due to thermal torques from the Sun (the YORP effect; Rubincam 2000; e.g., (54509) YORP, (1862) Apollo, (1620) Geographos, (3103) Eger; Lowry et al. 2007, Taylor et al. 2007; Kaasalainen et al. 2007; {\v D}urech et al. 2008, 2012); and asteroids were observed while \textbf{shedding} mass and even \textbf{breaking} apart (e.g. P/2010 A2, P/2013 P5, P/2013 R3; Jewitt et al. 2010, 2013, 2014).

	Indirect measurements also support our understanding of the rotational-fission mechanism: 1.) The lack of asteroids with $D>\sim100~m$ that rotate faster than 2.2~h per cycle suggests that such asteroids were slowed down by mass shedding or completely disintegrated into smaller $D<\sim100~m$ fragments (e.g., Richardson et al. 1998, Pravec \& Harris 2000, Warner et al. 2009, Jacobson et al. 2014a). 2.) The physical and dynamical parameters of binary asteroids and asteroid pairs were shown to match the expected outcome from a rotational-fission event - mainly a correlation between a fast rotation of the primary body to other parameters in the system (e.g., Pravec et al. 2006, Walsh et al. 2008, Pravec et al. 2010, Polishook et al. 2011, Fang \& Margot 2012, Jacobson et al. 2014b, Polishook 2014).

	This study was motivated by the idea that the fission event might reveal sub-surface material with a different spectral signature, and these would be detectable by rotationally-resolved observations of asteroid spectra. We focused our study on primary members of {\it asteroid pairs} that are believed to have formed recently (Vokrouhlick{\'y} \& Nesvorn{\'y} 2008) by a fission of a fast rotating asteroid (Pravec et al. 2010). An asteroid pair consists of two unbound objects with almost identical heliocentric orbital elements. Dynamical calculations have shown that the secondary members of each pair were in the Hill sphere of the larger member in the last 2 million years, indicating a single origin. Indeed, spectral measurements show that members of the same pair present the same spectral and color characteristics supporting a single origin of each pair (Moskovitz 2012, Duddy et al. 2012, 2013, Polishook et al. 2014, Wolters et al. 2014). Finding an area with a unique spectral signature on primary member of an asteroid pair could not only give clues to the way asteroids disrupt but also give us unique opportunity to get a glimpse of the sub-surface material of asteroids.

\subsection{Spectral variation on asteroids}
\label{sec:spectralVariation}
	The main reason to expect spectral heterogeneity on the young asteroid pairs is the {\it space weathering} mechanism that modifies the surface of atmosphere-less bodies by different agents such as solar radiation and micrometeorite bombardment (Chapman 2004). This mechanism is known for maturing the spectral signature of common taxonomical types such as ordinary chondrites by increasing their spectral slope and gradually erasing their $1 \mu m$ absorption band (Clark et al. 2002). Laboratory experiments were able to reproduce S-type reflectance spectra by laser irradiation of ordinary chondrite (e.g., Moroz et al. 1996, Sasaki et al. 2001, Brunetto et al. 2006). Space weathering effects on other asteroidal types (such as carbonaceous chondrites/C-complex, achondrite/V-type) are minor as inferred from astronomical (e.g., Pieters et al. 2000, Rivkin et al. 2002) and experimental (e.g., Vernazza et al. 2013, Brunetto et al. 2014) studies. Even though it is not clear how long does it take for the space weathering to mature a spectrum of an ordinary chondrite (studies present a wide range of possible timescales ranging from a $10^5$ to $10^8$ years; Vernazza et al. 2009, Willman et al. 2010), it seems from dynamic calculations of near-Earth asteroids that 100,000 years is a lower limit for an asteroid to present a non-weathered, ``fresh" material (Nesvorn{\'y} et al. 2010). {\it Asteroid pairs} are especially relevant in this context since they presumably disrupted in a range of $10^4$ to $2x10^6$ years, therefore, sub-surface ordinary chondrite revealed by their disruption might did not have the time to mature by space weathering. Furthermore, the fact that most asteroid pairs present ``non-fresh" spectral features (Polishook et al. 2014, Wolters et al. 2014) suggests that the possible fresh material on the site of the fission should be more noticeable.

	Spectral heterogeneity on asteroid surfaces is known to exist from in situ measurements collected by spacecrafts such as {\it Galileo} and {\it NEAR Shoemaker}. These found ``colorful" spots (areas with color differing than the average color) on craters located on asteroid (951) Gaspra (Veverka et al. 1996), and spectral variations on the surface of asteroid (433) Eros (Murchie \& Pieters 1996). However, measuring spectral variations from the Earth is quite challenging since the parameters of the asteroid spectrum are altered by the Earth's atmosphere (such as the spectral slope). For example, claims for spectral variability on the young asteroid, (832) Karin (Sasaki et al. 2004) were revoked by Chapman et al. (2007) and Vernazza et al. (2007). Currently, spectral heterogeneity is clearly observed from Earth on (4) Vesta, one of the largest asteroids in the main belt (Bobrovnikoff 1929, Gaffey 1997), and this was supported by space-based observations (Reddy et al. 2010, 2012). The large scale heterogeneity on Vesta is probably primordial and due to compositional differences across the crust and excavation of the crust to different depths. Smaller scale heterogeneity is probably due to impacts (Le Corre et al. 2013).

	Spectral heterogeneity is not limited to space weathering effects. Spectral heterogeneity was also measured in the near-IR range on near-Earth asteroids such as (101955) Bennu, (162173) 1999 JU3, (175706) 1996 FG3 and (285263) 1998 QE2 that do not present ordinary chondrite spectra. The heterogeneity in all these cases was measured at different epochs and it was suggested to be due to variations in viewing aspect (Moskovitz et al. in revision; Sanchez et al. 2012; Reddy et al. 2012b), compositional or grain-size effects (Binzel \& DeMeo 2013), thermal effects (Binzel et al. 2012), or a mere artificial cause, such as bad weather or calibration (Moskovitz et al. 2013). However, the real reason(s) for the spectral variation on these asteroids is not known.

\section{Observations and Reduction}
\label{sec:observations}

\subsection{Observed asteroids}
\label{sec:objects}

	Eleven asteroids were observed in this study. The candidate pair asteroids were taken from Pravec and Vokrouhlick{\'y} (2009) and Vokrouhlick{\'y} (2009). Excluding telescope availability, we limit target selection by a visible limiting magnitude of 18.5 and by a relatively short rotation period of $<6~h$ in order to complete the rotational coverage in a single night. An exception is the asteroid (54041), with a rotation period of about 18.86 hours. The physical information of these asteroids is summarized in Table~\ref{tab:PairsParam}.

	Eight of the asteroids were observed in near-IR range, two in the visible range, and one asteroid, (6070) {\it Rheinland}, was observed in both ranges.

\begin{deluxetable*}{lcccccccccc}
\tablecolumns{11}
\tablewidth{0pt}
\tablecaption{Pairs' physical properties and age}
\tablehead{
\colhead{Asteroid} &
\colhead{a} &
\colhead{$H_v$} &
\colhead{D} &
\colhead{Spin} &
\colhead{Amp} &
\colhead{Secondary} &
\colhead{${\Delta H}$} &
\colhead{$D _2$} &
\colhead{Age} &
\colhead{Taxonomy} \\
\colhead{}          &
\colhead{[AU]} &
\colhead{}          &
\colhead{[km]} &
\colhead{[hours]}          &
\colhead{[mag]}    &
\colhead{}          &
\colhead{[mag]}    &
\colhead{[km]}    &
\colhead{[kyrs]} &
\colhead{} \\
}
\startdata
 \hline
2110 Moore-Sitterly   & 2.20 & 13.2 & 6.5   & 3.3447  & 0.45 & 44612     & 2.3 & 2.3 & $>1600$      & S  \\
3749 Balam   & 2.24 & 13.1 & 6.8   & 2.8049  & 0.14 & 312497    & 4.4 & 0.9 & $280_{-25}^{+45}$    & Sq \\
4765 Wasserburg   & 1.95 & 13.7 & 3.7   & 3.625  & 0.56 & 350716    & 3.8 & 1.1 & $170_{-30}^{+430}$   & X/E \\
5026 Martes  & 2.38 & 13.8 & 9.6   & 4.4243  & 0.49 & 2005WW113 & 4   & 1.5 & $18\pm1$     & Ch \\
6070 Rheinland  & 2.39 & 13.7 & 5.2   & 4.2735  & 0.41 & 54827     & 1.6 & 2.5 & $17\pm0.5$   & Sq \\
10484 Hecht  & 2.32 & 13.8 & 3.8   & 5.508  & 0.21 & 44645     & 1   & 2.4 & $310_{-80}^{+210}$   & V \\
11842 Kap'bos & 2.25 & 13.9 & 4.7   & 3.6858  & 0.13 & 228747    & 2.9 & 1.5 & $>150$   & Sq/Q \\
17198 Gorjup & 2.28 & 14.9 & 3.0   & 3.243  & 0.13 & 229056    & 2.6 & 0.9 & $230_{-50}^{+120}$   & Sw \\
25884 2000SQ4 & 1.95 & 14.6 & 2.4   & 4.917  & 0.55 & 48527     & 1.5 & 1.2 & $420_{-100}^{+200}$  & X/E \\
54041 2000GQ113 & 2.32 & 14.5 & 2.8   & 18.86 & 0.23 & 220143    & 2   & 1.1 & $150_{-30}^{+470}$   & V \\
63440 2001MD30 & 1.94 & 15.2 & 1.8   & 3.2969  & 0.17 & 331933    & 2.2 & 0.7 & $33_{-4}^{+17}$      & X/E
\enddata
\tablenotetext{}{Source of data:}
\tablenotetext{-}{Semi-major axis, absolute magnitudes, and $\Delta H$ are from the MPC website, http://www.minorplanetcenter.net/.}
\tablenotetext{-}{Age and taxonomy are from Polishook et al. 2014, excluding data for 11842: age from Pravec et al. 2010, and taxonomy is by this study. The asteroids with ``X/E" taxonomy have a X-type like spectra; they were marked as E-type (a sub-group of the X-type; DeMeo et al. 2009) since their orbits lay within the Hungaria group that is rich with E-type asteroids.}
\tablenotetext{-}{Diameters were estimated from the absolute magnitude assuming an albedo value of 0.22 for S-complex asteroids, 0.36 for V-type, 0.43 for E-type and 0.058 for the Ch-type asteroid (Mainzer et al. 2011).}
\tablenotetext{-}{Rotation periods and amplitude were derived in this study (excluding data of 54041; its data was taken from Pravec et al. 2010). They match data from the literature derived by Pravec et al. 2010 (2110, 4765, 5026, 6070, 10484, 11842, 17198, 63440), Polishook 2011 (25884) and Polishook et al. 2011 (3749).}
\tablenotetext{-}{The secondary information is from Pravec and Vokrouhlick{\'y} (2009) and Vokrouhlick{\'y} (2009).}
\label{tab:PairsParam}
\end{deluxetable*}

\subsection{Infrared Spectroscopy}
\label{sec:infraredSpectroscopy}

	We conducted near-infrared spectroscopy (0.8 to 2.5 $\mu m$) using SpeX, an imager and spectrograph mounted on the 3-m telescope of NASA's InfraRed Telescope Facility (IRTF; Rayner et al. 2003). A long slit with a 0.8 arcsec width was used and the objects were shifted along it in an A-B-B-A sequence to allow the measurement of the background noise. The slit was aligned to the parallactic angle. Exposures of 120 seconds were taken for each image. Observations were limited to low air mass values to reduce chromatic refraction that can change the spectral slope. Almost all observations (excluding those of asteroid 4765) were taken at low phase angle of $<15^o$ in order to measure maximal area of the asteroids' surfaces as possible and to avoid any spectral change due to increasing phase angle ({\it phase reddening}; Sanchez et al. 2012). Solar analog stars were observed during the night to allow the correction of telluric lines and normalization by the Sun's spectrum. In addition, G2 to G5 stars, that were in close proximity to the asteroids, were also observed during the asteroid rotation to further calibrate each longitudinal segment of the asteroid's reflectance spectrum separately. Lamps and arcs images were also routinely taken to calibrate the CCD sensitivity map and the dispersion solutions, respectively. The observational details are listed in Table~\ref{tab:ObsCircumIrSpec}. An average reflectance spectrum for each asteroid, made out of these observations were published in a previous paper (Polishook et al. 2014).
	
	The reduction of the raw SpeX images follows the procedures outlined in Binzel et al. (2010) and DeMeo et al. (2009). This includes flat field correction, sky subtraction, manual aperture selection, background and trace determination, removal of outliers, and a wavelength calibration using arc images. A telluric correction routine was used to model and remove telluric lines. Each spectrum was divided by a standard solar analog to derive the relative reflectance of the asteroid (stars are listed in Table~\ref{tab:ObsCircumIrSpec}). The reflectance spectra of each asteroid were divided to represent segments of different rotational phases. The time span for each segment was set between 20 to 40 minutes (2 to 4 cycles of A-B-B-A sequences) to balance the need for sufficient S/N (longer exposures) and the desire to achieve the highest possible longitudinal resolution on the asteroids' surfaces (shorter exposures). The reflectance spectrum of each longitudinal segment was further divided by the reflectance spectrum of a G2/G5 star, observed circa the time and coordinates of the segment and normalized by the average reflectance spectrum of the star from the entire night. In this way temporal variations in the spectral slope of the asteroid's longitudinal segment, due to atmospheric instability, were corrected.

\begin{deluxetable*}{ccccccccccc}
\tablecolumns{11}
\tablewidth{0pt}
\tablecaption{Observational details - near-IR spectroscopy, IRAF 3m SpeX}
\tablehead{
\colhead{Asteroid} &
\colhead{longitudinal segment} &
\colhead{Date} &
\colhead{Total exposure time} &
\colhead{${\it N}$} &
\colhead{${\it R}$} &
\colhead{${\it \Delta}$} &
\colhead{${\it \alpha}$} &
\colhead{${\it V_{mag}}$} &
\colhead{Solar Analog} &
\colhead{Normalizing star}  \\
\colhead{}          &
\colhead{}          &
\colhead{}    &
\colhead{[minutes]}          &
\colhead{}          &
\colhead{[AU]} &
\colhead{[AU]} &
\colhead{[deg]} &
\colhead{}          &
\colhead{}          &
\colhead{}
}
\startdata
2110 & A & 2011 Oct 25 & 32 & 16 & 1.96 & 0.97 & 3.9 & 15.0 & L93-101 & SAO092992 \\
 & B & 2011 Oct 25 & 28 & 14 & 1.96 & 0.97 & 3.9 & 15.0 & L93-101 & SAO092992 \\
 & C & 2011 Oct 25 & 24 & 12 & 1.96 & 0.97 & 3.9 & 15.0 & L93-101 & SAO092992 \\
 & D & 2011 Oct 25 & 32 & 16 & 1.96 & 0.97 & 3.9 & 15.0 & L93-101 & SAO092992 \\
 & E & 2011 Oct 25 & 32 & 16 & 1.96 & 0.97 & 3.9 & 15.0 & L93-101 & SAO092992 \\
 & F & 2011 Oct 26 & 32 & 16 & 1.97 & 0.97 & 3.3 & 14.9 & L93-101 & SAO092992 \\
 & G & 2011 Oct 26 & 32 & 16 & 1.97 & 0.97 & 3.3 & 14.9 & L93-101 & SAO092992 \\
 & H & 2011 Oct 26 & 32 & 16 & 1.97 & 0.97 & 3.3 & 14.9 & L93-101 & SAO092992 \\
 & I & 2011 Oct 26 & 32 & 16 & 1.97 & 0.97 & 3.3 & 14.9 & L93-101 & SAO092992 \\
3749 & A & 2012 Jan 22 & 16 & 8 & 2.00 & 1.01 & 0.6 & 14.7 & L98-978 & SAO80041 \\
 & B & 2012 Jan 22 & 12 & 6 & 2.00 & 1.01 & 0.6 & 14.7 & L98-978 & SAO80041 \\
 & C & 2012 Jan 22 & 14 & 7 & 2.00 & 1.01 & 0.6 & 14.7 & L98-978 & SAO80041 \\
 & D & 2012 Jan 22 & 16 & 8 & 2.00 & 1.01 & 0.6 & 14.7 & L98-978 & SAO80041 \\
 & E & 2012 Jan 22 & 16 & 8 & 2.00 & 1.01 & 0.6 & 14.7 & L98-978 & SAO80041 \\
 & F & 2012 Jan 22 & 12 & 6 & 2.00 & 1.01 & 0.6 & 14.7 & L98-978 & SAO80041 \\
 & G & 2012 Jan 22 & 16 & 8 & 2.00 & 1.01 & 0.6 & 14.7 & L98-978 & SAO80041 \\
4765 & A & 2013 Jan 10 & 32 & 16 & 1.84 & 1.53 & 32.4 & 17.3 & L105-56 & HD119649 \\
 & B & 2013 Jan 10 & 32 & 16 & 1.84 & 1.53 & 32.4 & 17.3 & L105-56 & HD119649 \\
 & C & 2013 Jan 10 & 32 & 16 & 1.84 & 1.53 & 32.4 & 17.3 & L105-56 & HD119649 \\
 & D & 2013 Jan 11 & 32 & 16 & 1.84 & 1.52 & 32.4 & 17.3 & L105-56 & HD119649 \\
 & E & 2013 Jan 11 & 32 & 16 & 1.84 & 1.52 & 32.4 & 17.3 & L105-56 & HD119649 \\
 & F & 2013 Jan 11 & 24 & 12 & 1.84 & 1.52 & 32.4 & 17.3 & L105-56 & HD119649 \\
6070 & A & 2013 Oct 03 & 40 & 20 & 1.94 & 1.31 & 28.3 & 17.0 & Hya64 & --- \\
 & B & 2013 Oct 31 & 16 & 8 & 1.98 & 1.12 & 19.0 & 16.4 & L98-978 & --- \\
 & C & 2013 Dec 10 & 16 & 8 & 2.05 & 1.07 & 3.8 & 15.8 & Hya64 & --- \\
10484 & A & 2011 Oct 27 & 8 & 4 & 2.21 & 1.24 & 7.1 & 16.5 & L113-276 & SAO092217 \\
 & B & 2011 Oct 27 & 28 & 14 & 2.21 & 1.24 & 7.1 & 16.5 & L93-101 & SAO092217 \\
 & C & 2011 Oct 27 & 32 & 16 & 2.21 & 1.24 & 7.1 & 16.5 & L93-101 & SAO092217 \\
 & D & 2011 Oct 27 & 32 & 16 & 2.21 & 1.24 & 7.1 & 16.5 & L93-101 & SAO092217 \\
 & E & 2011 Oct 27 & 40 & 20 & 2.21 & 1.24 & 7.1 & 16.5 & L113-276 & SAO092217 \\
17198 & A & 2012 Oct 18 & 16 & 8 & 2.38 & 1.39 & 0.8 & 17.6 & L93-101 & SAO109993 \\
 & B & 2012 Oct 18 & 32 & 16 & 2.38 & 1.39 & 0.8 & 17.6 & L93-101 & SAO109993 \\
 & C & 2012 Oct 18 & 32 & 16 & 2.38 & 1.39 & 0.8 & 17.6 & L93-101 & SAO109993 \\
 & D & 2012 Oct 19 & 48 & 24 & 2.38 & 1.39 & 0.8 & 17.6 & L93-101 & SAO109993 \\
25884 & A & 2011 Oct 25 & 8 & 4 & 1.93 & 0.96 & 8.7 & 16.5 & L93-101 & SAO093282 \\
 & B & 2011 Oct 25 & 16 & 8 & 1.93 & 0.96 & 8.7 & 16.5 & L93-101 & SAO093282 \\
 & C & 2011 Oct 26 & 28 & 14 & 1.93 & 0.96 & 8.1 & 16.5 & Hya64 & SAO093282 \\
 & D & 2011 Oct 26 & 28 & 14 & 1.93 & 0.96 & 8.1 & 16.5 & Hya64 & SAO093282 \\
 & E & 2011 Oct 29 & 24 & 12 & 1.94 & 0.96 & 6.1 & 16.4 & Hya64 & SAO093282 \\
 & F & 2011 Oct 29 & 32 & 16 & 1.94 & 0.96 & 6.1 & 16.4 & Hya64 & SAO093282 \\
 & G & 2011 Oct 29 & 32 & 16 & 1.94 & 0.96 & 6.1 & 16.4 & Hya64 & SAO093282 \\
54041 & A & 2012 Nov 10 & 24 & 12 & 2.27 & 1.30 & 7.8 & 17.4 & Hya64 & --- \\
 & B & 2012 Dec 14 & 48 & 24 & 2.32 & 1.39 & 10.4 & 17.7 & Hya64 & HD23169 \\
 & C & 2012 Dec 17 & 28 & 14 & 2.32 & 1.41 & 11.7 & 17.8 & Hya64 & HD23169 \\
 & D & 2012 Dec 19 & 46 & 23 & 2.32 & 1.43 & 12.6 & 17.8 & Hya64 & HD23169 \\
63440 & A & 2012 Oct 19 & 32 & 16 & 1.81 & 0.84 & 9.6 & 16.7 & L93-101 & --- \\
 & B & 2012 Nov 09 & 26 & 13 & 1.80 & 0.83 & 10.9 & 16.8 & L93-101 & SAO109993 \\
 & C & 2012 Nov 09 & 32 & 16 & 1.80 & 0.83 & 10.9 & 16.8 & L93-101 & SAO109993 \\
 & D & 2012 Nov 09 & 36 & 18 & 1.80 & 0.83 & 10.9 & 16.8 & L93-101 & SAO109993 \\
 & E & 2012 Nov 10 & 32 & 16 & 1.80 & 0.84 & 11.5 & 16.8 & L93-101 & SAO109993
\enddata
\tablenotetext{}{Columns: asteroid's designation, longitudinal segment marker, date of observation, total exposure time, number of images, heliocentric and geocentric distances, phase angle, average magnitude (according MPC), solar analog and additional normalizing star.}
\label{tab:ObsCircumIrSpec}
\end{deluxetable*}

\subsection{Visible Spectroscopy}
\label{sec:visibleSpectroscopy}

	Additional observations were taken over the visible range with Las Campanas Observatory's 6.5m Magellan/Clay telescope using the LDSS3 instrument. The observations were performed on June 1, 2 and 3, 2012, with a 1.5''-wide and 4'-long slit and its VPH-All grism. These settings produced a useful spectral range of 0.44 to 0.94 $\mu m$ at a dispersion of 1.89 {\AA}/pixel. Exposures of 120 or 180 seconds were taken for each image at two different nod positions along the slit. Well-documented solar analogs (Bus \& Binzel 2002, DeMeo et al. 2009, Moskovitz et al. 2013) were observed throughout the nights for calibration purposes. The observational details are listed in Table~\ref{tab:ObsCircumVisSpec}. Reduction was performed in a standard way using IRAF and IDL routines. Arc lamp spectra of He, Ne and Ar were obtained to provide dispersion solutions and flat field calibration was performed by exposures of a white screen illuminated by a quartz lamp. Further details on the observation method and reduction process are described by Moskovitz et al. (2013). The reflectance spectra of each asteroid were divided to represent longitudinal segments of different rotational phases. The time span for each longitudinal segment was set between $\sim10$ to $\sim30$ minutes depending on the S/N. The obtained asteroid spectra of each longitudinal segment were divided by the solar analogs and normalized at 0.55 $\mu m$ to produce relative reflectance spectra. An average reflectance spectrum for each asteroid was constructed from the spectra of all segments.

\begin{deluxetable*}{cccccccccc}
\tablecolumns{10}
\tablewidth{0pt}
\tablecaption{Observational details - visible spectroscopy, Magellan 6.5m LDSS3}
\tablehead{
\colhead{Asteroid} &
\colhead{longitudinal segment} &
\colhead{Date} &
\colhead{Total exposure time} &
\colhead{${\it N}$} &
\colhead{${\it R}$} &
\colhead{${\it \Delta}$} &
\colhead{${\it \alpha}$} &
\colhead{${\it V_{mag}}$} &
\colhead{Solar Analog} \\
\colhead{}          &
\colhead{}          &
\colhead{}    &
\colhead{[minutes]}          &
\colhead{}          &
\colhead{[AU]} &
\colhead{[AU]} &
\colhead{[deg]} &
\colhead{}          &
\colhead{}
}
\startdata
5026 & A &  2012 Jun 01 & 9 & 3 & 2.42 & 1.72 & 20.8 & 17.9 & L102-1081 \\
 & B &  2012 Jun 01 & 15 & 5 & 2.42 & 1.72 & 21.1 & 17.9 & L102-1081 \\
 & C &  2012 Jun 02 & 12 & 4 & 2.42 & 1.73 & 21.1 & 17.9 & L102-1081 \\
 & D &  2012 Jun 02 & 15 & 5 & 2.42 & 1.73 & 21.1 & 17.9 & L102-1081 \\
 & E &  2012 Jun 03 & 12 & 4 & 2.41 & 1.74 & 21.3 & 18 & L105-56 \\
 & F &  2012 Jun 03 & 12 & 4 & 2.41 & 1.74 & 21.3 & 18 & L105-56 \\
 & G &  2012 Jun 03 & 12 & 4 & 2.41 & 1.74 & 21.3 & 18 & L105-56 \\
 & H &  2012 Jun 03 & 12 & 4 & 2.41 & 1.74 & 21.3 & 18 & L105-56 \\
 & I &  2012 Jun 03 & 12 & 4 & 2.41 & 1.74 & 21.3 & 18 & L105-56 \\
 & J &  2012 Jun 03 & 12 & 4 & 2.41 & 1.74 & 21.3 & 18 & L105-56 \\
6070 & A &  2012 Jun 01 & 8 & 4 & 2.64 & 1.63 & 2.8 & 17.2 & HD149182 \\
 & B &  2012 Jun 01 & 8 & 4 & 2.64 & 1.63 & 2.8 & 17.2 & HD149182 \\
 & C &  2012 Jun 01 & 10 & 5 & 2.64 & 1.63 & 2.8 & 17.2 & HD149182 \\
 & D &  2012 Jun 01 & 8 & 4 & 2.64 & 1.63 & 2.8 & 17.2 & HD149182 \\
 & E &  2012 Jun 01 & 12 & 6 & 2.64 & 1.63 & 2.8 & 17.2 & HD149182 \\
 & F &  2012 Jun 01 & 8 & 4 & 2.64 & 1.63 & 2.8 & 17.2 & HD149182 \\
 & G &  2012 Jun 02 & 8 & 4 & 2.64 & 1.63 & 3.2 & 17.2 & HD149182 \\
11842 & A &  2012 Jun 01 & 12 & 4 & 2.43 & 1.83 & 22.3 & 18.2 & L112-1333 \\
 & B &  2012 Jun 01 & 12 & 4 & 2.43 & 1.83 & 22.3 & 18.2 & L112-1333 \\
 & C &  2012 Jun 02 & 12 & 4 & 2.43 & 1.82 & 22.1 & 18.2 & L113-276 \\
 & D &  2012 Jun 02 & 18 & 6 & 2.43 & 1.82 & 22.1 & 18.2 & L113-276 \\
 & E &  2012 Jun 03 & 12 & 4 & 2.43 & 1.8 & 22 & 18.2 & L112-1333 \\
 & F &  2012 Jun 03 & 12 & 4 & 2.43 & 1.8 & 22 & 18.2 & L112-1333 \\
 & G &  2012 Jun 03 & 12 & 4 & 2.43 & 1.8 & 22 & 18.2 & L112-1333 \\
 & H &  2012 Jun 03 & 12 & 4 & 2.43 & 1.8 & 22 & 18.2 & L112-1333 \\
 & I &  2012 Jun 03 & 12 & 4 & 2.43 & 1.8 & 22 & 18.2 & L112-1333 \\
 & J &  2012 Jun 03 & 12 & 4 & 2.43 & 1.8 & 22 & 18.2 & L112-1333 \\
 & K &  2012 Jun 03 & 12 & 4 & 2.43 & 1.8 & 22 & 18.2 & L112-1333 \\
 & L &  2012 Jun 03 & 12 & 4 & 2.43 & 1.8 & 22 & 18.2 & L112-1333 \\
 & M &  2012 Jun 03 & 12 & 4 & 2.43 & 1.8 & 22 & 18.2 & L112-1333 \\
 & N &  2012 Jun 03 & 12 & 4 & 2.43 & 1.8 & 22 & 18.2 & L112-1333
\enddata
\tablenotetext{}{Columns: asteroid's designation, longitudinal segment marker, date of observation, total exposure time, number of images, heliocentric and geocentric distances, phase angle, average magnitude (according MPC) and solar analog.}
\label{tab:ObsCircumVisSpec}
\end{deluxetable*}

\subsection{Visible Photometry}
\label{sec:visibleRegime}
	In order to be able to state the rotational phase and the location on the asteroid that the spectral measurements were taken, we conducted photometric observations during the same apparition of the spectroscopic observations (in most cases at the same week). Asteroid 54041 was not observed since its rotation period is long ($\sim 18.86~hours$, Pravec et al. 2010) and we did not have enough observing time allocated. Observations were performed at the Wise Observatory in Israel using its 0.46m telescope (Brosch et al. 2008). The 0.46-m is equipped with a wide-field SBIG STL-6303E CCD (75'x55' with 3072x2048 pixels, 1.47'' per pixel, unbinned). Observations were performed without filters (``Clear"). To achieve a point-like FWHM for the moving targets (at a seeing value of $\sim$2.5 pixels), exposure times of $60-180$ seconds were used, all with an auto-guider. The observational circumstances are summarized in Table~\ref{tab:ObsCircumVisPhot}.

	The images were reduced in a standard way using bias and dark subtraction, and were divided by a normalized flatfield image. {\it IRAF}'s {\it phot} function was used for the photometry. Apertures with four-pixel radii were chosen to minimize photometric errors. The mean sky value was measured using an annulus of 10 pixels wide and inner radius of 10 pixels around the asteroid. The photometric values were calibrated to a differential magnitude level using local comparison stars measured on every image using the same method as the asteroid. The brightness of the comparison stars remained constant to $\pm$0.02 mag. A photometric shift was calculated for each image compared to a good reference image, using the local comparison stars. The asteroid data were corrected for light-travel time and the magnitudes were reduced to one AU distance from the Sun and the Earth (Bowell et al. 1989). Refer to Polishook and Brosch (2009) for detailed description of the photometric procedures of observation, reduction and calibration using the 0.46-m. Some of the observations were previously published by Polishook (2014, accepted for publication).
	
	The calibrated data for each asteroid was folded by their known rotation period (2110, 4765, 6070, 10484, 17198, 63440 - Pravec et al. 2010; 25884 - Polishook 2011; 3749 - Polishook et al. 2011) and fit with a second-order Fourier series (Harris et al. 1989). The measured photometric variabilities are in excellent match to the known rotation periods. The folded lightcurves are plotted on top of an ellipse, representing the rotational phases of the asteroid where the spectral observations took place (see Appendix Figures). Since most of the photometric measurements were taken at the same week of the spectral observations, and since the rotation periods are known to a high accuracy of seconds, we are confident that the times of the rotational phases and the respective spectra are correctly matched. We note that even though the spectral longitudinal segments are proceeding to the right, this is only to match the lightcurve and does not assume a preferred sense of rotation.

\begin{deluxetable*}{clcccccc}
\tablecolumns{8}
\tablewidth{0pt}
\tablecaption{Observational details - visible photometry, Wise 0.46m}
\tablehead{
\colhead{Asteroid} &
\colhead{Date} &
\colhead{Total exposure time} &
\colhead{${\it N}$} &
\colhead{${\it R}$} &
\colhead{${\it \Delta}$} &
\colhead{${\it \alpha}$} &
\colhead{${\it V_{mag}}$} \\
\colhead{}          &
\colhead{}          &
\colhead{[hours]}          &
\colhead{}    &
\colhead{[AU]} &
\colhead{[AU]} &
\colhead{[deg]} &
\colhead{}
}
\startdata
2110  & 2011 Oct 27 & 2.37 & 48 & 1.97 & 0.98 & 2.09 & 14.9 \\
  & 2011 Oct 28 & 2.82 & 56 & 1.97 & 0.98 & 1.62 & 14.8 \\
3749  & 2012 Jan 23 & 9.55 & 222 & 1.99 & 1.01 & 1.07 & 14.8 \\
4765  & 2013 Jan 11 & 3.98 & 61 & 1.84 & 1.52 & 32.39 & 17.3 \\
  & 2013 Jan 14 & 4.11 & 61 & 1.83 & 1.49 & 32.36 & 17.3 \\
  & 2013 Jan 15 & 3.27 & 50 & 1.83 & 1.48 & 32.35 & 17.2 \\
  & 2013 Jan 18 & 3.91 & 64 & 1.83 & 1.46 & 32.29 & 17.2 \\
5026  & 2012 Apr 16 & 2.98 & 45 & 2.54 & 1.54 & 3.91 & 17.1 \\
  & 2012 Apr 19 & 6.13 & 95 & 2.53 & 1.54 & 5.04 & 17.2 \\
6070  & 2012 May 25 & 5.42 & 55 & 2.65 & 1.64 & 0.65 & 17 \\
  & 2013 Oct 25 & 5.88 & 87 & 1.97 & 1.15 & 21.76 & 16.5 \\
  & 2013 Oct 28 & 2.03 & 33 & 1.97 & 1.13 & 20.47 & 16.4 \\
  & 2014 Jan 03 & 0.87 & 14 & 2.1 & 1.22 & 15.9 & 16.6 \\
  & 2014 Jan 06 & 7.86 & 92 & 2.11 & 1.25 & 17.18 & 16.7 \\
10484  & 2011 Oct 31 & 6.33 & 86 & 2.21 & 1.26 & 9.42 & 16.6 \\
11842  & 2012 Jul 27 & 3.96 & 50 & 2.39 & 1.38 & 2.95 & 16.8 \\
17198  & 2012 Oct 16 & 6.95 & 71 & 2.38 & 1.39 & 1.3 & 17.7 \\
25884  & 2011 Oct 28 & 4.83 & 69 & 1.94 & 0.95 & 6.19 & 16.4 \\
63440  & 2012 Nov 06 & 8.45 & 79 & 1.8 & 0.83 & 9.72 & 16.7 \\
  & 2012 Nov 08 & 3.7 & 53 & 1.8 & 0.83 & 10.86 & 16.8 \\
  & 2012 Dec 06 & 5.06 & 64 & 1.78 & 0.98 & 25.23 & 17.6
\enddata
\tablenotetext{}{Columns: asteroid's designation, date of observation, total exposure time, number of images, heliocentric and geocentric distances, phase angle, average magnitude (according MPC).}
\label{tab:ObsCircumVisPhot}
\end{deluxetable*}

\section{Analysis}
\label{sec:analysis}

\begin{figure}
\centerline{\includegraphics[width=10cm]{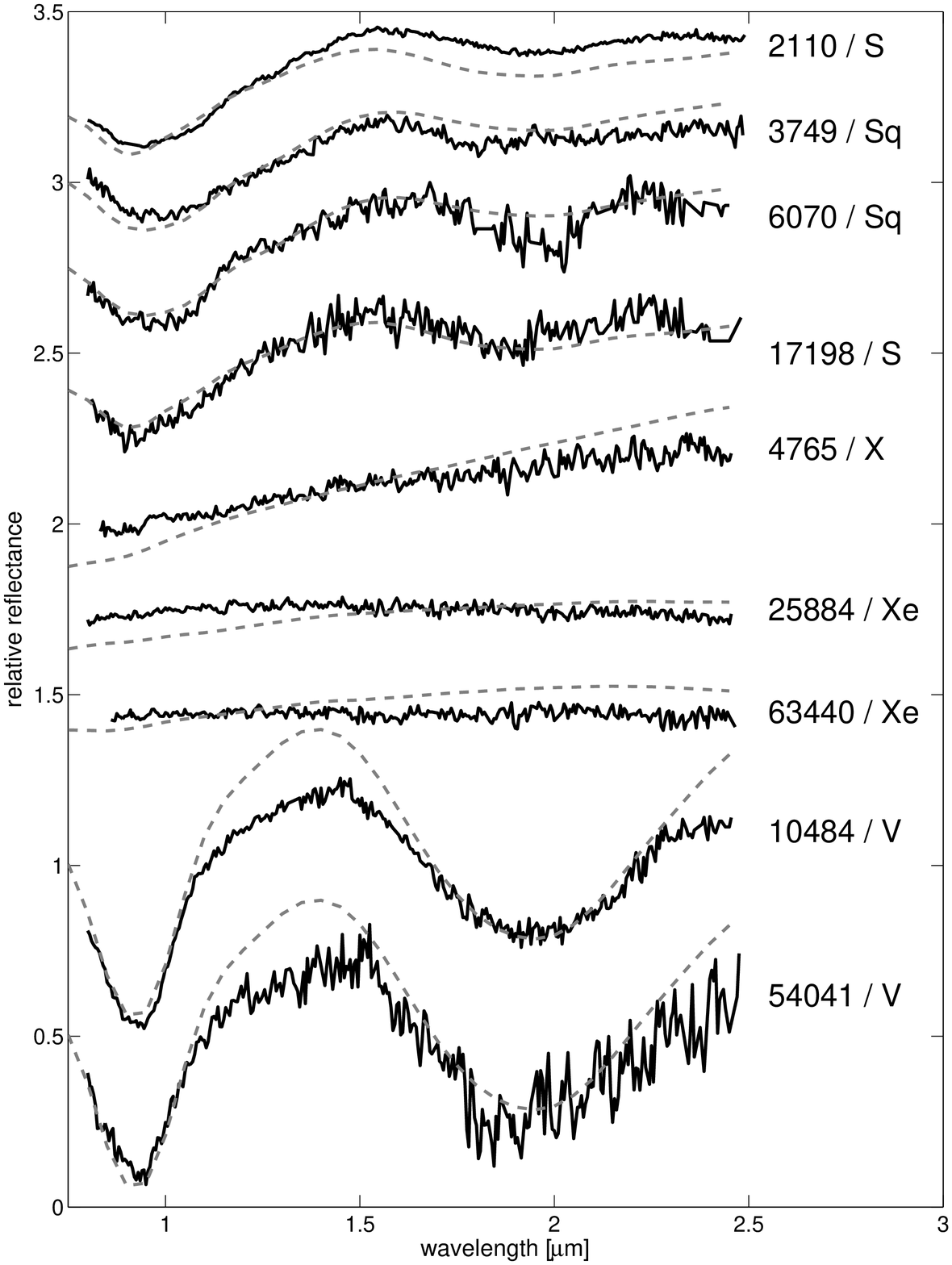}}
\caption{The mean near-infrared spectra of nine studied asteroids (solid black line). Taxonomic representatives are marked on top each spectrum (grey dash-line). Asteroids' names and taxonomy letter are written on the right. The spectra were shifted on the Y-axis for clarity.
\label{fig:PairsSpectra}}
\end{figure}

\begin{figure}
\centerline{\includegraphics[width=10cm]{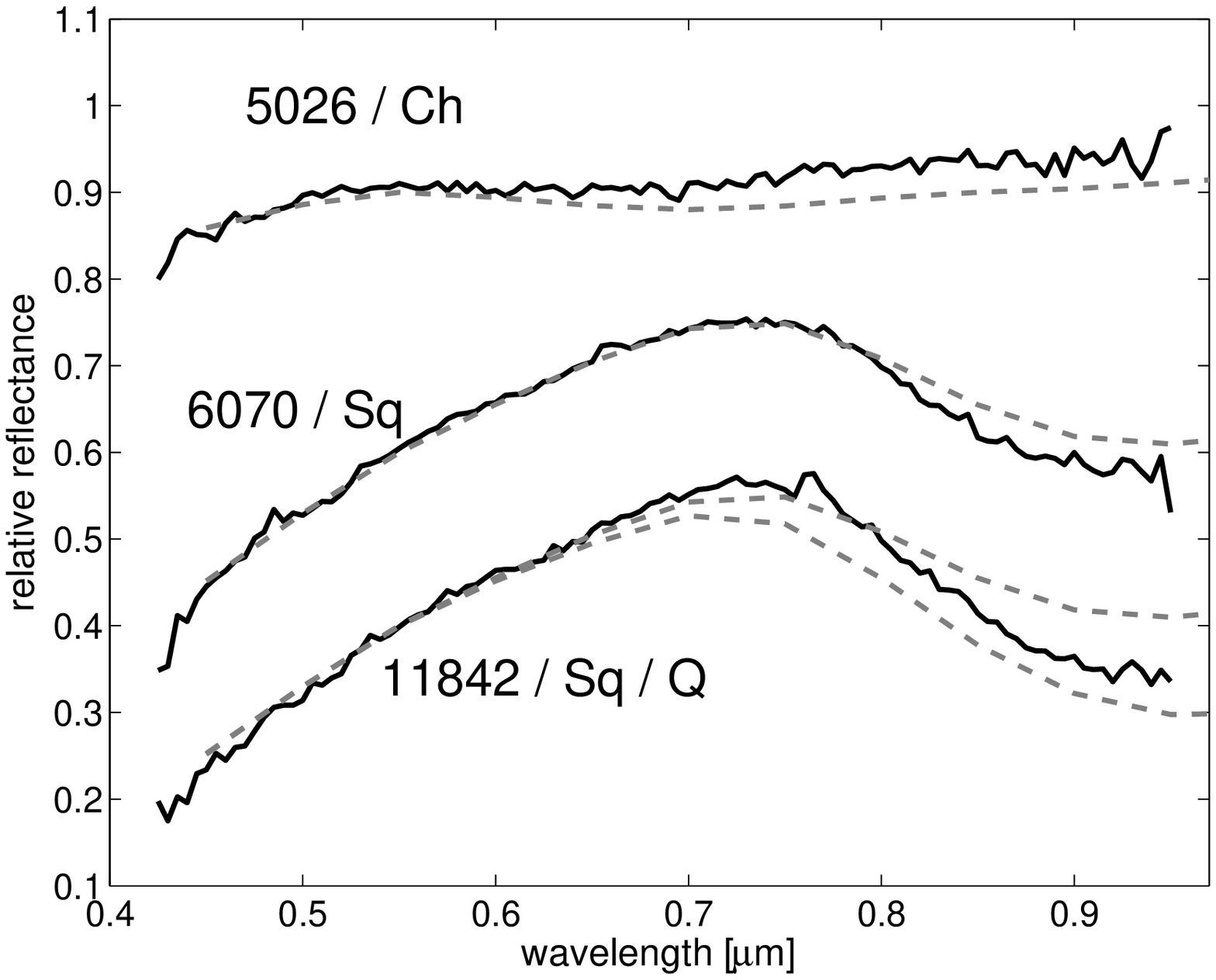}}
\caption{The mean visible spectra of three studied asteroids (solid black line). Taxonomic representatives are marked on top each spectrum (grey dash-line). Asteroids' names and taxonomy letter are written on the right. The spectra were shifted on the Y-axis for clarity.
\label{fig:PairsVisSpectra}}
\end{figure}

	The reflectance spectra of the asteroids, averaged from the spectra of all longitudinal segments, are presented in Fig.~\ref{fig:PairsSpectra} and \ref{fig:PairsVisSpectra}. Five of the spectra are of the S-complex taxonomy, three are of the flat X-complex, two have V-type spectra and one is a Ch-type\footnote{10 of these reflectance spectra were previously published by Polishook et al. (2014).}. The reflectance spectra of each longitudinal segment, normalized by the average spectrum of the asteroid are presented in Appendix A with the rotational phases at which they were taken compared to the folded lightcurve of the asteroids.
	
	The reflectance spectrum of each longitudinal segment was normalized by the average reflectance spectrum to reveal any potential variation in the spectrum. Here we refer to these as the {\bf segment's normalized reflectance spectrum}. In most cases the variation is subtle and within the noise of the measurement, especially at wavelengths with telluric lines that were not corrected properly, or at the edges where the sensitivity of the instrument decrease. In a few cases where a variation is visible, a real spectral variation was rejected since 1) there was another longitudinal segment, representing the same rotational phase but taken at a different time that did not display the same spectral variation; 2) the reflectance spectrum of the longitudinal segment was not normalized by data from a nearby G2/G5 star since one was not accessible. These cases are summarized in the captions of the figures on Appendix A.

\begin{figure}
\centerline{\includegraphics[width=10cm]{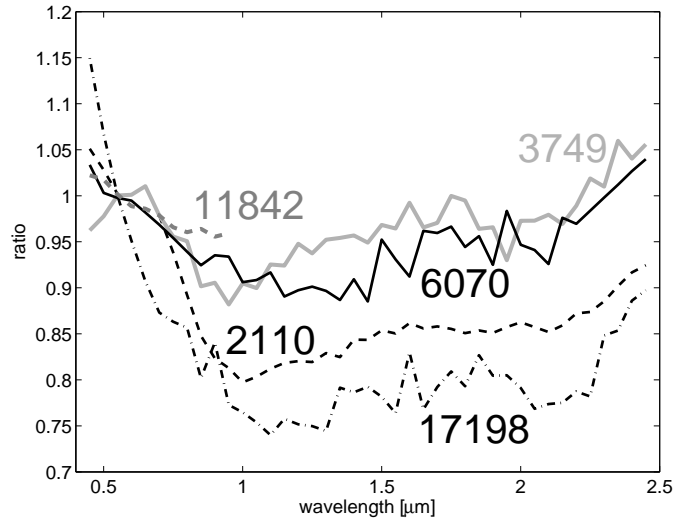}}
\caption{The expected variation in the spectra of the five ordinary chondrite asteroids in our sample, presented as the ratio between an average Q-type to the average spectra of these asteroids. Objects are from top to bottom: 11842 (Sq-type, grey dashed line), 6070 (Sq-type, black line), 3749 (Sq-type, grey line), 2110 (S-type, black dashed line) and 17198 (S-type, black dotted line). Note that the average spectra include the visible range that allows normalizing them to unity at 0.55 $\mu m$. The average spectra with the visible range is from Polishook et al. (2014). The change from unity for each of the asteroids is larger than the measurements' noise, therefore, a segment with a Q-type spectrum would have been recognized in our measurements.
\label{fig:QtypeNormalizedByPairsSpec}}
\end{figure}

	Examining the spectral variation only on the S-complex asteroids, we present how a fresh, Q-type spectrum will look like on each of the five S-complex asteroids in our sample (2110, 3749, 6070, 11842, 17198) after normalizing it by the average reflectance spectrum of each asteroid (Fig.~\ref{fig:QtypeNormalizedByPairsSpec}). The change from unity for each of the asteroids is larger than the measurements' noise (compare to the figures in Appendix A and to the {\it Noise} column in Tables~\ref{tab:spotSizesIR} and \ref{tab:spotSizesVis}) for all of the asteroids excluding some segments of 11842. Therefore, we are certain that any variation in the spectrum due to the existence of a fresh material on the surface with a Q-type spectrum, would have been detected above the noise of our measurements. Some subtle spectral variations measured on asteroid 11842 are rejected since these variations were not re-measured in the same rotational phase at a second cycle. To conclude, {\bf we could not find a single case of significant, repeatable, convincing, spectral variation}.

	The same null result is derived when the segments that represent the narrow side of the asteroids are compared to the segments that represent the wider side of the asteroids as indicated by the lightcurve. Such a comparison is relevant since according to numerical studies the fission should occur on the narrow side of the object (Richardson et al. 1998, S{\'a}nchez and Scheeres 2012).

	Not detecting spectral variation due to ``colorful" spots on the asteroid surface does not necessarily mean they do not exist. We reject the idea that such spots could be located on one of the asteroid poles since the rotational-fission should happen in a place of maximal angular speed, on the long axis of the asteroid, on its equator and far from the poles. We know that we did not observe the asteroids with our line of sight aligned along their spin axes since the amplitudes of their light curves, as measured at multiple epochs, support low obliquity rotation states (Pravec et al. 2010, Vokrouhlick{\'y} et al. 2011, Polishook 2014, accepted for publication).
	
	Therefore, we use our null-result in order to constrain the maximal area of a potential ``colorful" spot. We chose a conservative approach and assume an extreme case where the entire disc of the asteroid that is facing the observer, contributes light evenly, even though Lambert's scattering law says that the luminous intensity is directly proportional to the cosine of the angle between the line of sight and the surface normal. By overlooking Lambert's scattering law we place an upper limit on the extent of any ``colorful" spots.

	For each longitudinal segment normalized reflectance spectrum we calculate the maximal area of a non-detected spot. A spot will not be detected if its contribution to the spectrum of the longitudinal segment is smaller than the noise. We define the noise $N$ as the standard deviation of the longitudinal segment's normalized reflectance spectrum. The noise of our data ranges from $0.5\%$ to $12\%$ with a mean value of $\sim3\%$. The entire area visible to the observer is the entire disc of the asteroid plus the spherical lune that appear during the exposure time starting at $T_1$ and ending at $T_2$\footnote{$(T_2-T_1)/P$ does not equal zero since significant observing time was needed to obtain sufficient signal-to-noise. If the observing time was much shorter than the rotation period of the asteroid, then a high-resolution map could have been made to reduce potential ÒspotÓ sizes.}. For simplicity we assume the asteroid is a sphere and calculate its surface accordingly. This assumption probably also increases the maximal area of the ``colorful" spots since the site of the fission most probably happened on the narrower side of the asteroid that presents a smaller disc than a disc of an equivalent-size sphere. We take into account the phase angle of the asteroid during the observation and use only the illuminated fraction of the disc $f_i$. The size of the maximal area $A_{max}$ of a non-detected spot is

\begin{equation}
A_{max} = (2\pi R^2f_i + 2\pi\frac{T_2-T_1}{P} 2R^2)\times N
\label{eq:maxSpotArea}
\end{equation}

where $P$ is the rotation period and $R$ is the radius of the asteroid. We define an effective diameter $D_{max}$ for the area $A_{max}$ as

\begin{equation}
D_{max} = 2\sqrt{A_{max}/\pi}.
\label{eq:maxSpotDiam}
\end{equation}

The calculation assumes that the asteroids' topography is flat, the edges of the potential spots are clearly defined and the material inside the spot is completely fresh and not partially fresh. Even though these assumptions are probably oversimplified, this analysis method could give a first order of magnitude of the size of a potential ``spot" and allow us to compare this value to the size of the secondary member of the pair that was supposedly ejected from the location of the potential spot. Assuming the secondary member was torn from the asteroid at the site of fission, they should have similar sizes (Scheeres 2007, Pravec et al. 2010, Jacobson \& Scheeres 2011). If the diameter of a secondary object is larger than the maximal effective diameter of the spot, than we can conclude that the fission event did not result in the exposure of material with a different reflectance spectrum.

	The results, displayed in Tables~\ref{tab:spotSizesIR} and \ref{tab:spotSizesVis}, show that the secondary members of asteroids 2110, 6070, 10484, 25884 and 63440 are larger by approximately a factor of two than a potential, undetected, ``colorful" spot (factors for specific asteroids: 2110 - $2.1$, 6070 - $2.8$, 10484 - $2.0$, 25884 - $1.7$, 63440 - $1.4$; Tables~\ref{tab:spotSizesIR} and \ref{tab:spotSizesVis}). Therefore, we can say for certain for these objects that on the area where the fission happened the material has a reflectance spectrum similar to other areas on the asteroids.

\begin{deluxetable*}{ccccccc}
\tablecolumns{7}
\tablewidth{0pt}
\tablecaption{Maximal effective diameters of potential spots from near-IR spectroscopy}
\tablehead{
\colhead{Asteroid} &
\colhead{$D_2$} &
\colhead{longitudinal segment} &
\colhead{$f_i$} &
\colhead{Noise} &
\colhead{$D_{max}$} &
\colhead{$A_{max}$} \\
\colhead{}          &
\colhead{[km]}          &
\colhead{}          &
\colhead{}          &
\colhead{}          &
\colhead{[km]}          &
\colhead{[$km^2$]}
}
\startdata
2110 & 2.3 & A & 1 & 0.015 & 1.3 & 1.3 \\
 &  & B & 1 & 0.006 & 0.8 & 0.5 \\
 &  & C & 1 & 0.005 & 0.7 & 0.4 \\
 &  & D & 1 & 0.009 & 1.0 & 0.8 \\
 &  & E & 1 & 0.005 & 0.8 & 0.5 \\
 &  & F & 1 & 0.01 & 1.1 & 0.9 \\
 &  & G & 1 & 0.006 & 0.9 & 0.6 \\
 &  & H & 1 & 0.008 & 1.0 & 0.7 \\
 &  & I & 1 & 0.009 & 1.0 & 0.8 \\
3749 & 0.9 & A & 1 & 0.014 & 1.3 & 1.3 \\
 &  & B & 1 & 0.02 & 1.5 & 1.7 \\
 &  & D & 1 & 0.028 & 1.8 & 2.5 \\
 &  & E & 1 & 0.014 & 1.2 & 1.2 \\
 &  & F & 1 & 0.02 & 1.5 & 1.8 \\
 &  & G & 1 & 0.015 & 1.3 & 1.4 \\
4765 & 1.1 & A & 0.92 & 0.045 & 1.3 & 1.2 \\
 &  & B & 0.92 & 0.045 & 1.3 & 1.2 \\
 &  & C & 0.92 & 0.042 & 1.2 & 1.2 \\
 &  & D & 0.92 & 0.022 & 0.9 & 0.6 \\
 &  & E & 0.92 & 0.024 & 0.9 & 0.7 \\
 &  & F & 0.92 & 0.024 & 0.9 & 0.6 \\
6070 & 2.5 & A & 0.94 & 0.073 & 2.3 & 4.2 \\
 &  & B & 0.97 & 0.041 & 1.5 & 1.8 \\
 &  & C & 1 & 0.033 & 1.4 & 1.5 \\
10484 & 2.4 & A & 1 & 0.036 & 1.1 & 0.9 \\
 &  & B & 1 & 0.019 & 0.8 & 0.5 \\
 &  & C & 1 & 0.025 & 0.9 & 0.7 \\
 &  & D & 1 & 0.024 & 1.0 & 0.7 \\
 &  & E & 1 & 0.037 & 1.2 & 1.1 \\
17198 & 0.9 & A & 1 & 0.044 & 1.0 & 0.7 \\
 &  & B & 1 & 0.07 & 1.4 & 1.5 \\
 &  & C & 1 & 0.046 & 1.1 & 0.9 \\
 &  & D & 1 & 0.046 & 1.3 & 1.3 \\
25884 & 1.2 & A & 0.99 & 0.028 & 0.6 & 0.3 \\
 &  & B & 0.99 & 0.019 & 0.5 & 0.2 \\
 &  & C & 0.99 & 0.023 & 0.6 & 0.3 \\
 &  & D & 0.99 & 0.024 & 0.6 & 0.3 \\
 &  & E & 1 & 0.036 & 0.7 & 0.4 \\
 &  & F & 1 & 0.021 & 0.6 & 0.3 \\
 &  & G & 1 & 0.036 & 0.7 & 0.4 \\
54041 & 1.1 & A & 1 & 0.118 & 1.4 & 1.5 \\
 &  & B & 0.99 & 0.065 & 1.1 & 0.9 \\
 &  & C & 0.99 & 0.125 & 1.4 & 1.6 \\
 &  & D & 0.99 & 0.067 & 1.1 & 0.9 \\
63440 & 0.7 & A & 0.99 & 0.025 & 0.5 & 0.2 \\
 &  & B & 0.99 & 0.031 & 0.5 & 0.2 \\
 &  & C & 0.99 & 0.019 & 0.4 & 0.1 \\
 &  & D & 0.99 & 0.027 & 0.5 & 0.2 \\
 &  & E & 0.99 & 0.021 & 0.4 & 0.1
\enddata
\tablenotetext{}{Column legend:}
\tablenotetext{}{Asteroid, diameter of the secondary asteroid, longitudinal segment marker, fraction of illuminated hemisphere, the standard deviation of the longitudinal segment's normalized reflectance spectrum, maximal diameter of an un-detected ``spot", maximal surface of an un-detected ``spot".}
\label{tab:spotSizesIR}
\end{deluxetable*}

\begin{deluxetable*}{ccccccc}
\tablecolumns{7}
\tablewidth{0pt}
\tablecaption{Maximal effective diameters of potential spots from visible spectroscopy}
\tablehead{
\colhead{Asteroid} &
\colhead{$D_2$} &
\colhead{longitudinal segment} &
\colhead{$f_i$} &
\colhead{Noise} &
\colhead{$D_{max}$} &
\colhead{$A_{max}$} \\
\colhead{}          &
\colhead{[km]}          &
\colhead{}          &
\colhead{}          &
\colhead{}          &
\colhead{[km]}          &
\colhead{[$km^2$]}
}
\startdata
5026 & 1.5 & A & 0.97 & 0.047 & 3.0 & 7.2 \\
     &     & B & 0.97 & 0.025 & 2.3 & 4.1 \\
     &     & C & 0.97 & 0.018 & 1.9 & 2.8 \\
     &     & D & 0.97 & 0.022 & 2.1 & 3.6 \\
     &     & E & 0.97 & 0.015 & 1.7 & 2.3 \\
     &     & F & 0.97 & 0.018 & 1.9 & 2.8 \\
     &     & G & 0.97 & 0.021 & 2.1 & 3.4 \\
     &     & H & 0.97 & 0.016 & 1.8 & 2.6 \\
     &     & I & 0.97 & 0.012 & 1.6 & 2.0 \\
     &     & J & 0.97 & .026 & 2.3 & 4.2 \\
6070	&	2.5	&	A	&	1	&	0.014	&	0.9	&	0.7	 \\
	&		&	B	&	1	&	0.008	&	0.7	&	0.4	 \\
	&		&	C	&	1	&	0.006	&	0.6	&	0.3	 \\
	&		&	D	&	1	&	0.01	&	0.8	&	0.5	 \\
	&		&	E	&	1	&	0.013	&	0.9	&	0.7	 \\
	&		&	F	&	1	&	0.01	&	0.8	&	0.5	 \\
	&		&	G	&	1	&	0.01	&	0.8	&	0.5	 \\
11842 & 1.5 & A & 0.96 & 0.015 & 1.7 & 2.4 \\
 &  & B & 0.96 & 0.016 & 1.8 & 2.5 \\
 &  & C & 0.96 & 0.011 & 1.4 & 1.6 \\
 &  & D & 0.96 & 0.023 & 2.2 & 3.9 \\
 &  & E & 0.96 & 0.03 & 2.4 & 4.7 \\
 &  & F & 0.96 & 0.015 & 1.7 & 2.3 \\
 &  & G & 0.96 & 0.016 & 1.8 & 2.5 \\
 &  & H & 0.96 & 0.017 & 1.9 & 2.7 \\
 &  & I & 0.96 & 0.014 & 1.6 & 2.1 \\
 &  & J & 0.96 & 0.018 & 1.9 & 2.8 \\
 &  & K & 0.96 & 0.014 & 1.7 & 2.2 \\
 &  & L & 0.96 & 0.023 & 2.2 & 3.7 \\
 &  & M & 0.96 & 0.024 & 2.2 & 3.8 \\
 &  & N & 0.96 & 0.026 & 2.3 & 4.0
\enddata
\tablenotetext{}{Column legend:}
\tablenotetext{}{Asteroid, diameter of the secondary asteroid, longitudinal segment marker, fraction of illuminated hemisphere, the standard deviation of the longitudinal segment's normalized reflectance spectrum, maximal diameter of an un-detected ``spot", maximal surface of an un-detected ``spot".}
\label{tab:spotSizesVis}
\end{deluxetable*}

\section{Discussion and Conclusions}
\label{sec:conclusions}

	We spectrally observed 11 asteroids in pairs in the near-IR and visible wavelength range as they rotate. Photometric observations were also conducted to derive the asteroid lightcurves and to determine the rotational phases of the spectral observations. We do not detect any rotational spectral variations within the signal-to-noise of our measurements, which allows us to tightly constrain the extent of any existing surface heterogeneity and show that for five asteroids a surface heterogeneity must be smaller than the size of the secondary object.

This can be explained in two ways: 1.) the material exposed from within the asteroid has the same spectrum as the material on its surface, or, 2.) a material with different spectrum does exist but it was covered again by dust and debris from the surface that were released by the fission event, and re-accumulate on the surface.

	While there is no reason to believe that the asteroid interior is compositionally different than the surface (asteroids smaller than $\sim20$ km are too small to have experienced differentiation; Moskovitz \& Gaidos 2011. All of the asteroids studied here are smaller than 20 km), we do expect spectral differences in the case of ordinary chondrite asteroids that can present distinctive weathered and fresh reflectance spectra\footnote{The surfaces of non-ordinary chondrite asteroids might also become weathered with time; however, this process seems to be very subtle (Vernazza et al. 2013, Brunetto et al. 2014). Therefore, we focus on the ordinary chondrite asteroids 2110 and 6070.}. Since asteroids 2110 and 6070 are of the S-complex taxonomy (i.e. they are ordinary chondrite asteroids; DeMeo et al. 2009), the fact that a ``colorful" spot with a different spectral signature was not detected supports the second scenario of re-accumulation of a mixture of weathered and fresh dust and debris that covers the expected `fresh" spot. This is especially evident for the asteroid 6070 that has an age ($\sim17~kyrs$) shorter than any published timescale of the space weathering mechanism (e.g., Vernazza et al. 2009, Nesvorn{\'y} et al. 2010). The fact that 6070's secondary member, 54827, presents a non-weathered, Q-type, spectrum while the spectrum of 6070 is more weathered, Sq-type (see Fig. 10 in Polishook et al. 2014) further demonstrates that the site of fission that should have the spectrum of the secondary member, has being covered by dust and debris that hides its true nature. We should note that the small-sized secondary member continues to disrupt due to torques applied on it by the larger primary member, exposing more sub-surface, fresh material (Jacobson \& Scheeres 2011, Polishook et al. 2014).

	Alternative scenarios fail to explain the lack of spectral variation on 6070. For example, an electrostatic migration of dust due to collisions or the YORP effect can be ruled out since the timescales for these mechanisms are several orders of magnitude longer than the short dynamical age of 6070 (Marzari et al. 2011, Rozitis \& Green 2013). In another model the disruption takes so long, that as the asteroid stretches the exposed fresh material becomes weathered before the asteroid is completely broken. However, numerical models show that the disruption of an asteroid due to fast rotation occurs in a matter of days, much shorter than the timescale of space weathering (Jacobson and Scheeres 2011, Sanchez and Scheeres 2012).

	Even though only the results for asteroid 6070 are conclusive we do not find any 	reason to believe that this interpretation is irrelevant for other asteroids in pairs unless they were formed by a different mechanism. The size of the ejected secondary object is probably an important parameter determining the amount of involved settling dust, but all of the objects reported here have a secondary member in the size range of 1 km and higher.
	
	The idea that the rotational-fission mechanism is followed with the spread of dust and debris is also supported by theoretical models (Richardson et al. 1998, S{\'a}nchez and Scheeres 2012) and is consistent with the recently discovered group of ``active" asteroids (Jewitt 2011): in three cases, that were observed since 2010 by HST, a sub-km size object with an asteroid-like orbit in the main belt was seen developing coma (P/2010 A2; Jewitt et al. 2010), forming multiple tails (P/2013 P5; Jewitt et al. 2013), and even disintegrating completely into smaller fragments (P/2013 R3; Jewitt et al. 2014). These three cases involved a large amount of dust production that was estimated by Jewitt et al. (2014) to be equivalent to a $\sim35~m$ radius sphere. If these cases of ``active" asteroids were formed by the rotational-fission mechanism, then this reported amount of dust produced could perhaps explain why at the area of fission fresh ordinary chondrite-like material was not observed. Therefore, we conclude that the constraints applied by our observations support the linkage between the group of ``active" asteroids to the rotational-fission mechanism, binary asteroids and asteroid pairs.

\acknowledgments

	We thank Pierre Vernazza and an anonymous referee for their useful comments. DP is grateful to the AXA research fund for their generous postdoctoral fellowship. NM acknowledges support from the Carnegie Institution of Washington, Department of Terrestrial Magnetism where the visible wavelength spectral observations were performed. NM also acknowledges support from the NSF Astronomy and Astrophysics Postdoctoral Program. FED acknowledges support for this work provided by NASA under Grant No. NNX12AL26G issued through the Planetary Astronomy Program and through the Hubble Fellowship grant HST-HF-51319.01-A, awarded by the Space Telescope Science Institute, which is operated by the Association of Universities for Research in Astronomy, Inc., for NASA, under contract NAS 5-26555.
	
	Observations reported here were obtained at the Infrared Telescope Facility, which is operated by the University of Hawaii under Cooperative Agreement NCC 5-538 with NASA, Science Mission Directorate, Planetary Astronomy Program. This paper also includes data gathered with the 6.5 meter Magellan Telescopes located at Las Campanas Observatory, Chile and from the Wise Observatory in Israel. Some of this material is based upon work supported by the National Science Foundation under Grant No. 0907766.

\begin{figure*}
\centerline{\includegraphics[width=17cm]{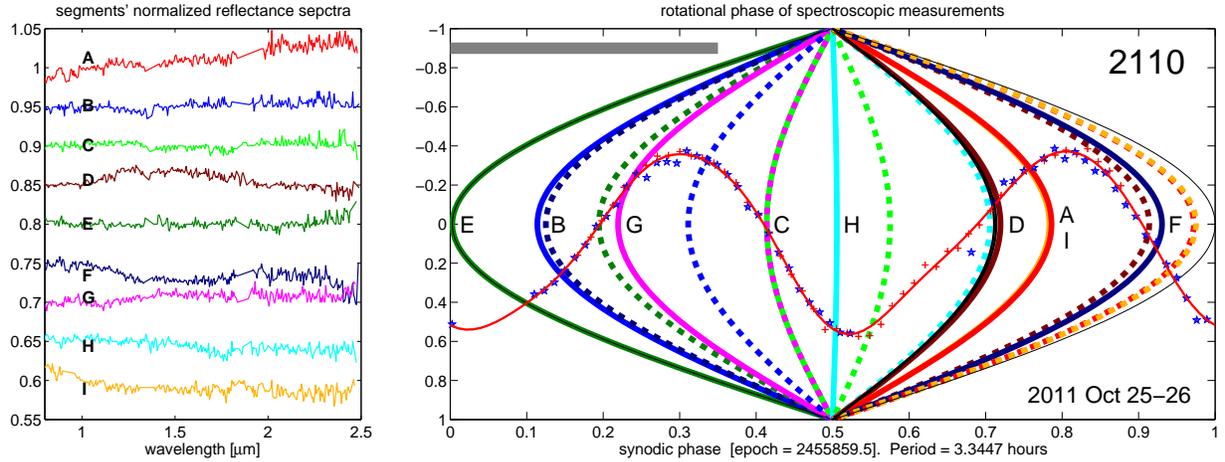}}
\caption{The normalized reflectance spectra of each longitudinal segment of asteroid (2110) {\it Moore-Sitterly} (left) and the times each longitudinal segment was taken marked on an ellipsoid (right). A segment's start is marked with a solid line and its end is marked with a dash line. The letters that appear after every segment beginning match those that appear on the normalized reflectance spectra, as do the colors. The folded lightcurve that was observed very close to the time the spectra were collected, is displayed on the right, and its phasing and epoch time match those of the spectra. The amplitude of the lightcurve is scaled to fit the mid-section of the graph and does not reflect the real photometric amplitude. A longitude that was not observed at sub-Earth location is marked with black. The grey line at the top left corner represent the diameter of the secondary member relative to the diameter of the primary member (values appear in Table~\ref{tab:PairsParam}). Pay attention that segments $A$ and $I$ were observed almost at the same time, therefore, the different slope of their normalized reflectance spectra is probably due to an atmospheric instability.
\label{fig:2110_SpecVari}}
\end{figure*}

\begin{figure*}
\centerline{\includegraphics[width=17cm]{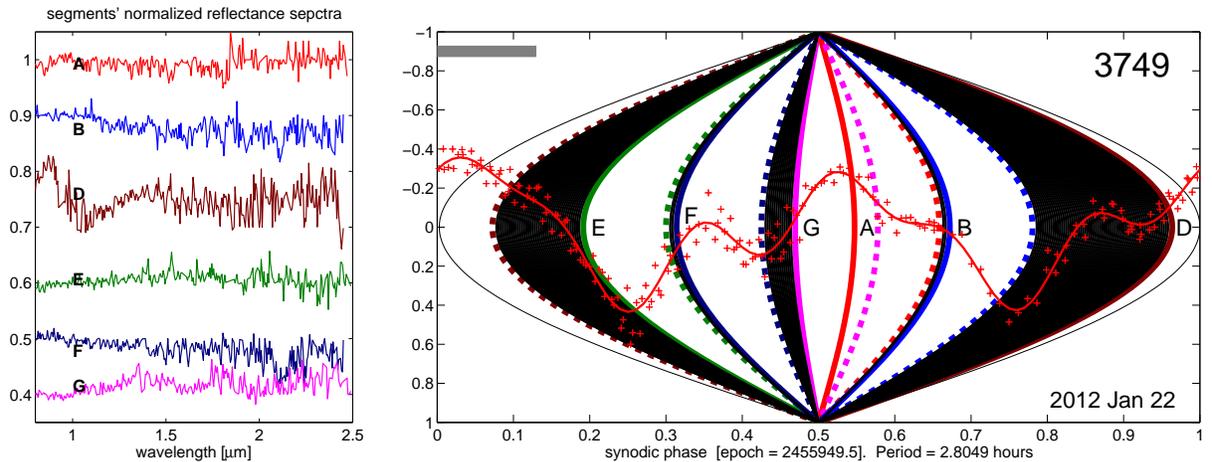}}
\caption{Same as Fig.~\ref{fig:2110_SpecVari} but for asteroid (3749) {\it Balam}. Segment $D$ displays a unique parameters of the $1\micron$ absorption band compared to the other segments. However, the spectrum of segment $D$ might have been effected by clouds that completely corrupted the spectrum of segment $C$ that is not displayed for this reason. Therefore, most likely this is an artificial spectral feature.
\label{fig:3749_SpecVari}}
\end{figure*}

\begin{figure*}
\centerline{\includegraphics[width=17cm]{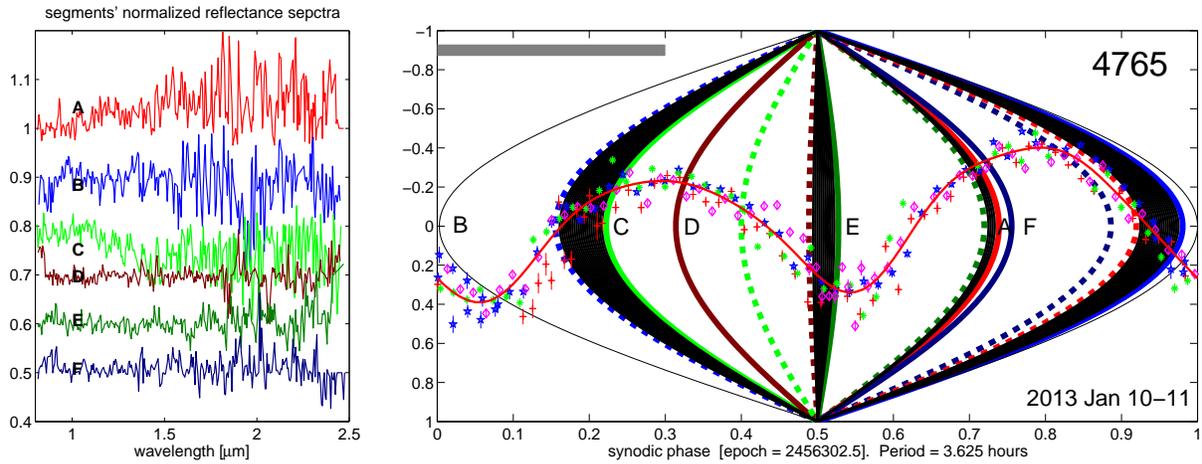}}
\caption{Same as Fig.~\ref{fig:2110_SpecVari} but for asteroid (4765) {\it Wasserburg}. Segment $C$ displays a lower spectral slope compared to the other segments. However, spectrum $D$ that was taken $\sim 30^o$ off segment $C$ does not display a low spectral slope. If a ``colorful" spot exists on part of segment $C$ that was not covered by segment $D$, than its effective diameter is $\sim1~km$.
\label{fig:4765_SpecVari}}
\end{figure*}

\begin{figure*}
\centerline{\includegraphics[width=17cm]{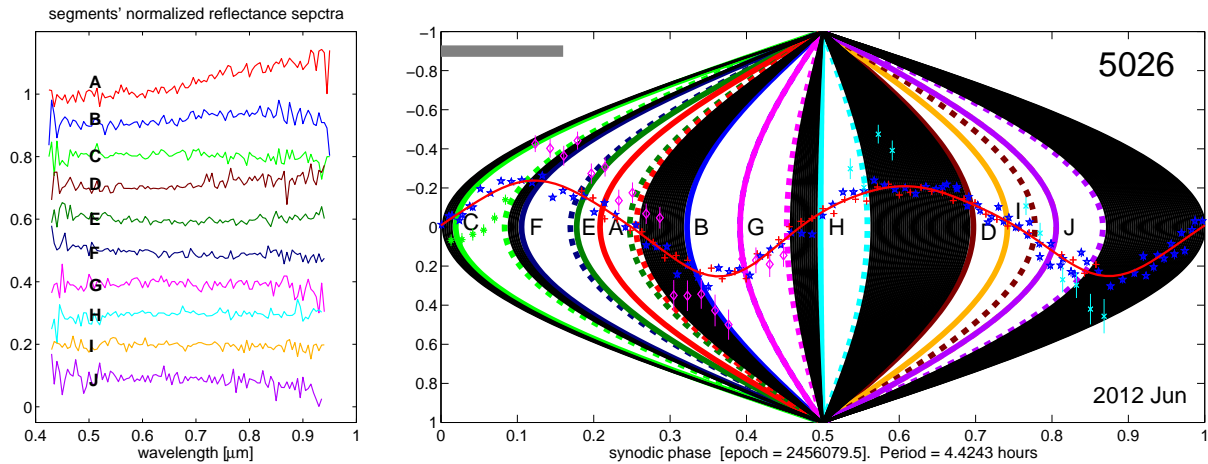}}
\caption{Same as Fig.~\ref{fig:2110_SpecVari} but for asteroid (5026) {\it Martes} at the visible range. Segment $A$ displays a higher spectral slope compared to the other segments. However, spectrum $E$ was taken at almost the same phase two nights later and it does not display the same spectral slope. Segment $J$ displays a lower spectral slope compared to the other segments. However, spectrum $I$ was taken at almost the same phase and it does not display the same spectral slope.
\label{fig:5026_SpecVari}}
\end{figure*}

\begin{figure*}
\centerline{\includegraphics[width=17cm]{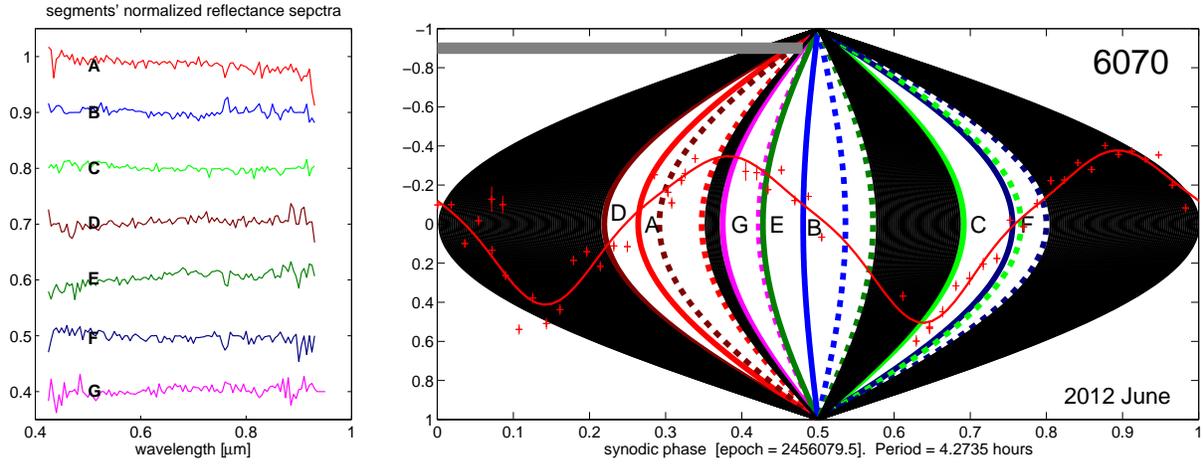}}
\caption{Same as Fig.~\ref{fig:2110_SpecVari} but for asteroid (6070) {\it Rheinland} at the visible range. Segment $A$ displays a lower spectral slope compared to the other segments. However, spectrum $D$ was taken at almost the same phase and it does not display the same spectral slope. Segment $E$ displays a higher spectral slope compared to the other segments. However, spectra $B$ and $G$ were taken at almost the same phase do not display the same spectral slope. Segment $F$ displays a lower spectral slope compared to the other segments. However, spectrum $C$ was taken at almost the same phase and it does not display the same spectral slope.
\label{fig:6070_VisSpecVari}}
\end{figure*}

\begin{figure*}
\centerline{\includegraphics[width=17cm]{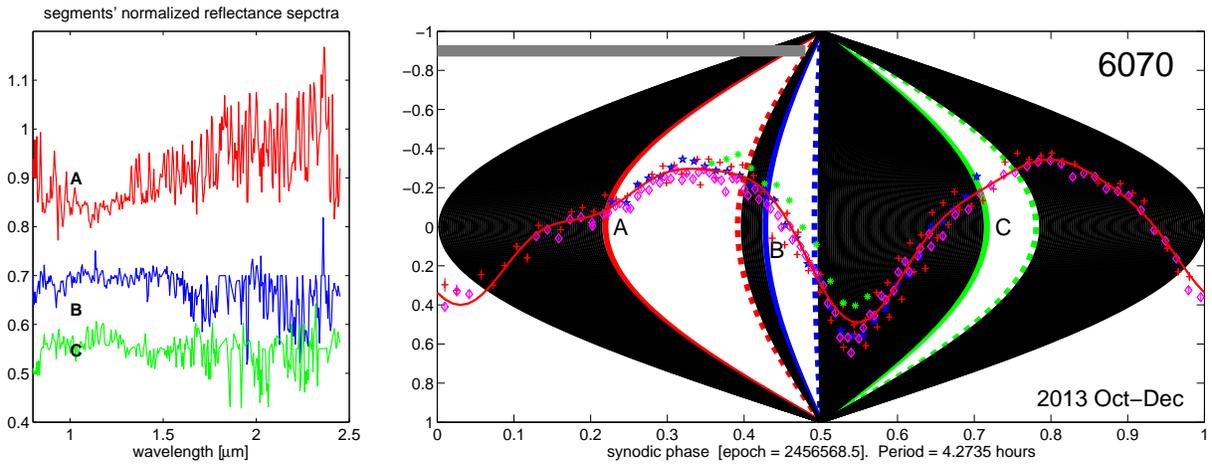}}
\caption{Same as Fig.~\ref{fig:2110_SpecVari} but for asteroid (6070) {\it Rheinland} at the near-IR range. We did not combined the visible and near-IR spectra and plotted them together since they were taken at different years (2012 and 2013) and the uncertainty on the period does not allow to securely combine visible and near-IR spectra of the same segments. Pay attention that the epoch is not the same. However, the lightcurve shape allows us to compare the rotational phases for the segments in the two spectral ranges. Segment $A$ of the IR spectra of asteroid 6070 displays a higher spectral slope compared to the other segments. However, the three reflectance spectra of this asteroid were not normalized by a nearby G2/G5 star, therefore, the slope difference might be due to atmospheric instability. Furthermore, 6070 does no present any spectral variation in the visible range, where it was observed more frequently.
\label{fig:6070_IrSpecVari}}
\end{figure*}

\begin{figure*}
\centerline{\includegraphics[width=17cm]{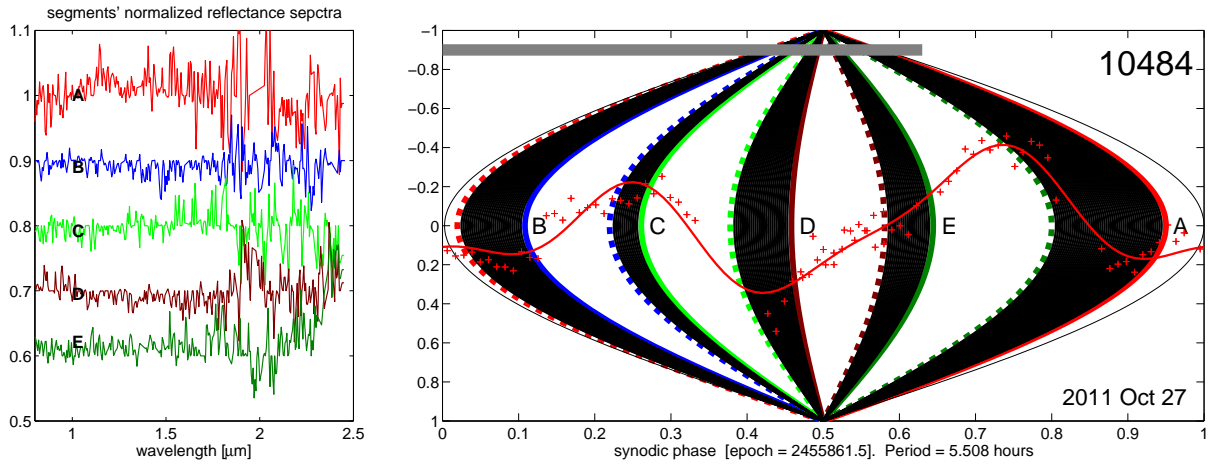}}
\caption{Same as Fig.~\ref{fig:2110_SpecVari} but for asteroid (10484) {\it Hecht}.
\label{fig:10484_SpecVari}}
\end{figure*}

\begin{figure*}
\centerline{\includegraphics[width=17cm]{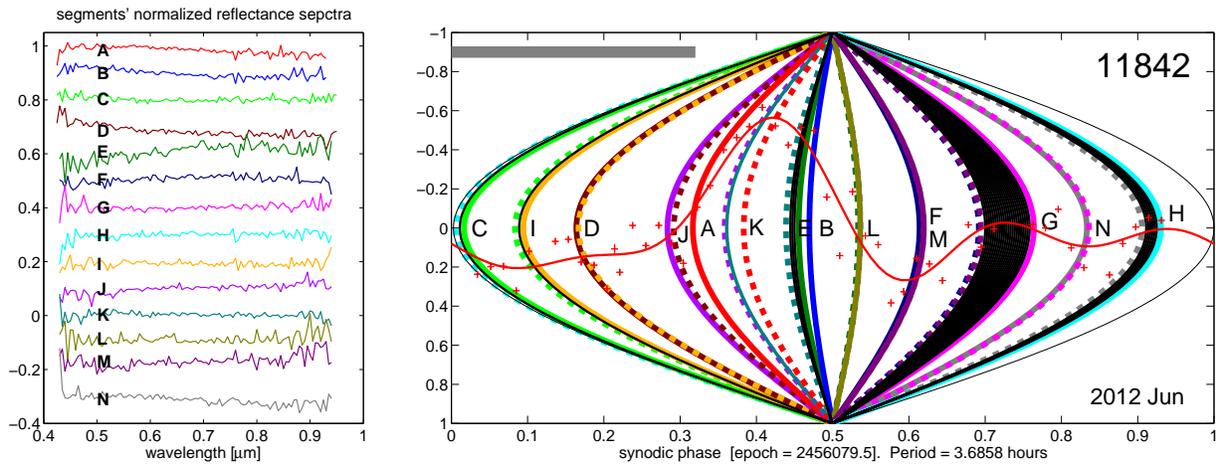}}
\caption{Same as Fig.~\ref{fig:2110_SpecVari} but for asteroid (11842) {\it Kap'bos} at the visible range. Segments $A$, $B$, $D$ and $N$ present significant lower spectral slopes compared to the other segments, that match a slope of a Q-type spectrum normalized by the average spectrum of 11842. However, these slopes do not repeat themselves in other cycles, suggesting these variations are due to atmospheric instability.
\label{fig:11842_VisSpecVari}}
\end{figure*}

\begin{figure*}
\centerline{\includegraphics[width=17cm]{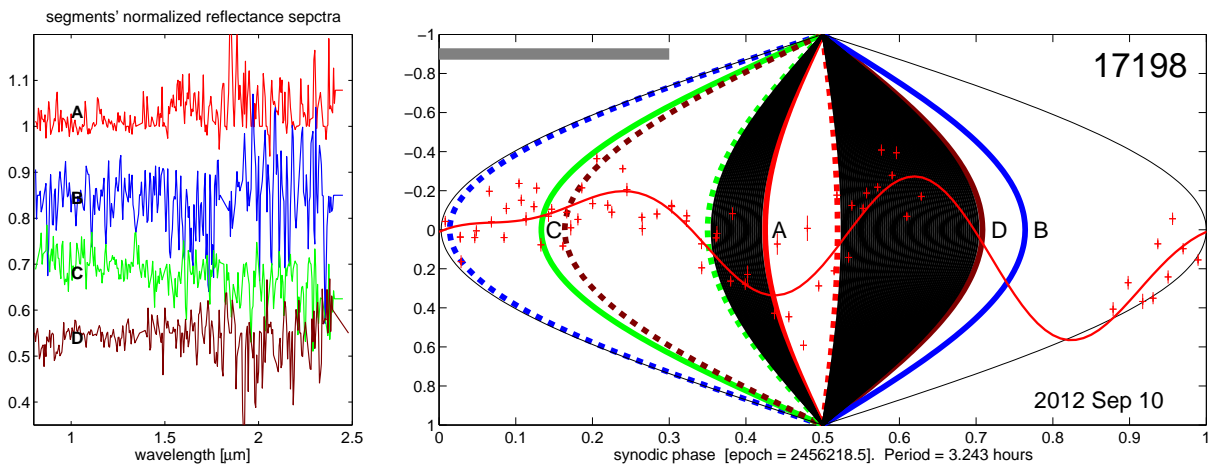}}
\caption{Same as Fig.~\ref{fig:2110_SpecVari} but for asteroid (17198) {\it Gorjup}.
\label{fig:17198_SpecVari}}
\end{figure*}

\begin{figure*}
\centerline{\includegraphics[width=17cm]{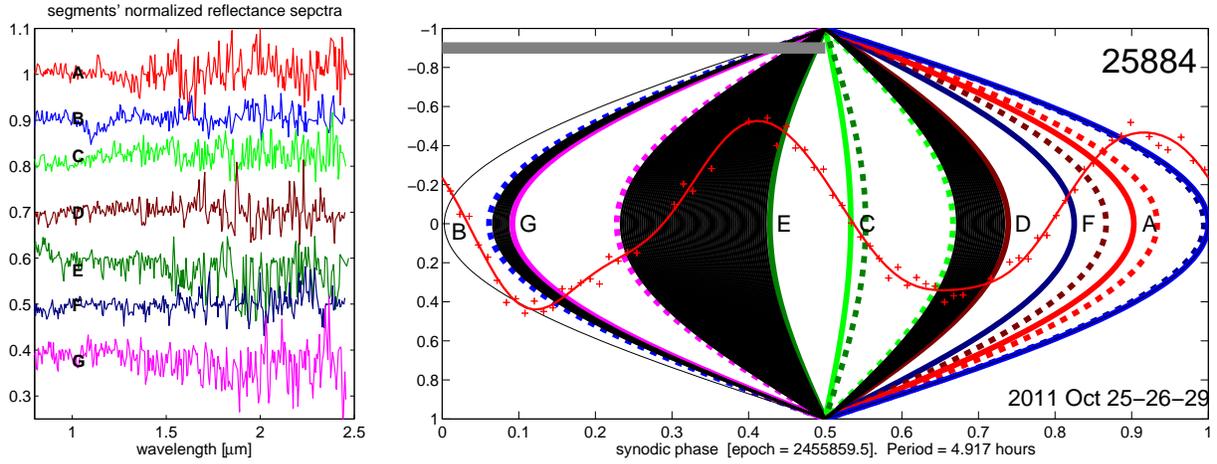}}
\caption{Same as Fig.~\ref{fig:2110_SpecVari} but for asteroid (25884) {\it 2000SQ4}.
\label{fig:25884_SpecVari}}
\end{figure*}

\begin{figure*}
\centerline{\includegraphics[width=17cm]{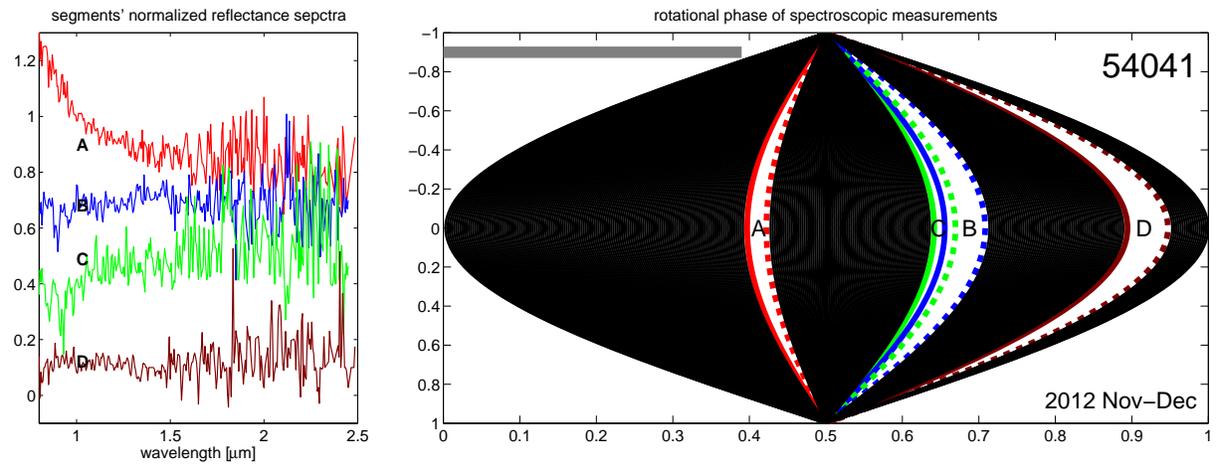}}
\caption{Same as Fig.~\ref{fig:2110_SpecVari} but for asteroid (54041) {\it 2000GQ113}. Segment $A$ displays a significant lower spectral slope compared to the other segments. However, this reflectance spectrum of 54041 is the only segment of this asteroid that was not normalized by a nearby G2/G5 star. Therefore, the slope difference might be due to atmospheric instability. Since the rotation period of 54041 is long (18.86 hours) we did not observe it photometrically.
\label{fig:54041_SpecVari}}
\end{figure*}

\begin{figure*}
\centerline{\includegraphics[width=17cm]{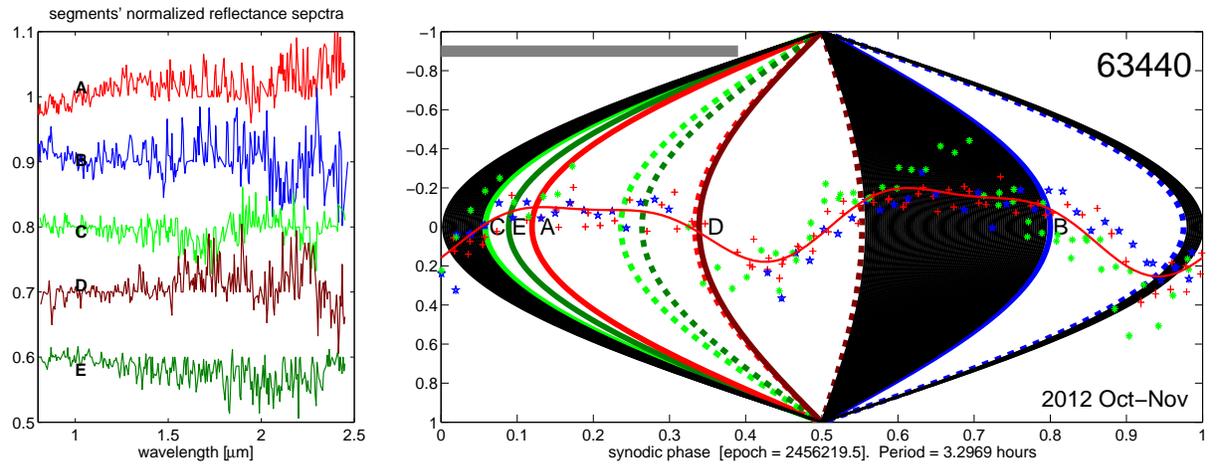}}
\caption{Same as Fig.~\ref{fig:2110_SpecVari} but for asteroid (63440) {\it 2001MD30}. Segment $A$ displays a slightly higher spectral slope compared to the other segments. However, the $A$ segment almost completely overlaps with the $E$ segment that presents a flat normalized reflectance spectrum.
\label{fig:63440_SpecVari}}
\end{figure*}

\appendix
\label{sec:ApendixA}
The reflectance spectra of each longitudinal segment, normalized by the average spectrum of the asteroid with the rotational phases at which they were taken compared to the folded lightcurve of the asteroids (Fig.~\ref{fig:2110_SpecVari}$-$~\ref{fig:63440_SpecVari}).

\end{document}